\documentclass[aps,twocolumn,epsfig,graphics,showpacs,floatfix,mathbbm]{revtex4}

\usepackage{amsmath,amsfonts,amssymb,graphics,graphicx,epsfig,color,times,bbm}

\newcommand{\nn}{\mathbbm{N}}

\newtheorem{observation}{Observation}

\begin{document}

\title{Minimal resources for linear optical one-way computing}

\author{K.\ Kieling, D.\ Gross, and J.\ Eisert}

\affiliation{
1 Blackett Laboratory, 
Imperial College London,
Prince Consort Road, 
London SW7 2BW, UK\\
2 Institute for Mathematical Sciences, Imperial College London,
Prince's Gardens, 
London SW7 2PE, UK
}
\date\today


\begin{abstract}
We address the question of how many 
maximally entangled photon pairs are needed in order to 
build up cluster states for quantum computing using 
the toolbox of linear optics. As the 
needed gates in dual-rail encoding are
necessarily probabilistic with known optimal 
success probability, this question amounts 
to finding the optimal strategy 
for building up cluster states,
from the perspective of classical control. 
We develop a notion of classical strategies, 
and present rigorous statements on the ultimate 
maximal and minimal use of resources of the globally 
optimal strategy. We find that this 
strategy -- being also the most robust with
respect to decoherence -- gives rise to an advantage 
of already more than an order of magnitude in the number of 
maximally entangled pairs when building chains with an 
expected length of $L=40$, compared to other legitimate 
strategies. For two-dimensional cluster states,
we present a first scheme achieving the 
optimal quadratic asymptotic scaling.
This analysis shows that the choice of 
appropriate classical control
leads to a very significant 
reduction in resource consumption.
\end{abstract}

\maketitle

To actually experimentally realize a fully-fletched
universal quantum computer constitutes a 
 tremendous challenge. Among the promising
 candidates for possible architectures
 are the ones entirely
 relying on optical systems. 
 State manipulation can
 then be realized using  
 sources of single photons
 or entangled pairs, arrays of linear optical 
 elements, and photon detectors 
 \cite{KLM,PC,Reznik,Nielsen04,Terry,Rev}. 
 Some of the advantages of such an approach
 are obvious: accurate state manipulation
 is available using linear optical elements, and photons
 are comparably robust with respect to decoherence.
 In turn, there is a price to pay when avoiding the 
exploitation of any physical non-linearities and
 effectively realizing them via measurements:
 due to the small success probability of elementary
 gates \cite{SP,SP2}, a very significant overhead in 
 optical elements and additional photons
is required to render the overall protocol near-deterministic.
 
Consequently, one of the primary goals of theoretical work towards the
realization of a linear optical quantum computer is to find ways to
reduce the necessary overhead in resources.  
For the seminal scheme of
Ref.\ \cite{KLM}, this overhead can not be reduced 
by simply building
better elementary sign shift gates \cite{SP}.
Schemes based on the model of one 
way computation \cite{Oneway} 
point towards a reduction of resource consumption
by orders of magnitude \cite{Nielsen04,Terry},
a perspective that has attracted considerable 
interest in recent research
\cite{Reznik,Nielsen04,Terry,BK,Kieling05a}. 
This development reminds of an inverse
`Moore's law' of the known minimally required
resources for linear
optical computing as a function of time.
The central ingredient
to these realizations are {\it cluster states} 
\cite{Oneway} 
or {\it graph states} \cite{GS} which
can be built up from maximally entangled 
photon pairs ($4$-qubit
cluster states have already been experimentally 
prepared \cite{Exp}).
Fusion gates of type\nobreakdash-I and~II 
have been applied to the task of creating cluster
states \cite{Terry,Ralph05}, deriving from parity 
check gates \cite{PC} and partial Bell projections.
However, these gates are inherently 
{\it probabilistic}, in that
in each step the experiment can either succeed or fail
with the outcome being known.

In fact, it is not difficult to show that the maximal
probability of success of a quantum gate realizing
a fusion of two dual-rail encoded linear cluster 
states is $p_s=1/2$, by relating this to the optimal success
probability of a Bell measurement with linear optics
\cite{Lut,Spec}, see also Ref.\ \cite{longpaper}.
When preparing linear cluster states from EPR
pairs, the only freedom we have for improvement
is to identify the optimal classical strategy for 
fusing cluster state pieces. 
As the possible patterns of failure and
success increase exponentially,
an overwhelming wealth of 
situations can potentially occur. Deciding
how to {\it optimally} react to any of these situations 
constitutes a very hard problem indeed, but may 
have tremendous implications on the
amount of resources 
needed. A similar situation occurs when 
preparing two-dimensional (2-D) cluster states. 

In this work we will address the latter question; i.e., 
what is the
optimal strategy to cope with the probabilistic nature of fusion gates
in constructing one and two-dimensional cluster states? While previous research has more strongly
focused on saving resources 
by devising ingenious ways of
implementing {\it quantum} gates, 
it is found in the present paper
that choosing an optimal 
{\it classical control strategy} can cut the
needed entanglement
by further orders of magnitude. In this
way, we can also bound the resources that any
scheme within the above mentioned set of rules
would require.

We begin the specific investigation with the one-dimensional case.
Linear cluster states can be pictured as {\it chains} of qubits,
characterized by their length $l$ given in the number of edges. 
Maximally entangled qubit pairs
(EPR pairs) 
correspond to chains with
a single edge.  By a {\it configuration} we mean a set of chains of
specific individual lengths.
Type-I fusion \cite{Terry} allows for operations involving end qubits
of two pieces (lengths $l_1$ and $l_2$), resulting on success
($p_{s}=1/2$) in a single piece of length $l_1+l_2$ or on failure
($p_{f}=1/2$) in two pieces of length $l_1-1$ and $l_2-1$. The
process starts with a collection of EPR pairs and ends when only a
single piece is left. 
A {\it strategy} decides which chains to fuse given a configuration.
It is assessed by the {\it expected length} of the final cluster.
The vast majority of strategies allow for no simple description and can
be specified solely by a `lookup table' listing all configurations
with the respective proposed action. Since the number of
configurations scales as 
$O(N^{1/2} \exp(\pi (2N/3)^{1/2}))$ \cite{dim}
as a function of the total number of edges $N$, 
a single strategy is already an
extremely complex object and any form of brute 
force optimization is completely out of reach.

However, there is one simple strategy which might reasonably be
conjectured to be optimal. Indeed, we face a probabilistic process,
and we lose entangled resources on average.  Hence, it seems
advantageous to quickly build up long clusters by always fusing the
largest available pieces together. 
This strategy we call {\sc Greed}:
\begin{itemize}
\item {\it {\sc Greed}: Always fuse the largest available 
pieces.}
\end{itemize}
In turn, one can also be conservative and always fuse the smallest
available pieces. This apparently inferior strategy, dubbed {\sc
Modesty},  will not deliver long chains in early steps. 
\begin{itemize}
\item {\it {\sc Modesty}: Always fuse the smallest available 
pieces.}
\end{itemize}
Quite surprisingly, it will turn out that not only is {\sc Modesty}
vastly more effective than {\sc Greed}, but even extremely 
close to the globally optimal strategy.

Let us further formalize these notions. A (pure)
configuration 
consisting of $n_i$ pieces of length $l_i$,
$i=1,..., c$, 
will be denoted as $C:=\{l_1^{(n_1)}, \dots,
l_c^{(n_c)}\}$.  The {\it total number of edges} is given by
$N(C):=\sum_i n_i l_i$ and $C_{N}:=\{C|N(C)\leq N\}$ is the
{\it configuration space} for $N \in \nn$.  A {\it mixed
configuration} is a probability distribution $p$ defined on the
elements of $C_{N}$. The {\it expected total length} of a mixed
configuration is $\langle L \rangle(p) := \sum_{C} p(C)  N(C)$.
Strategies act naturally as stochastic matrices
\cite{VK}
on mixed configurations by acting on every
pure configuration in its support independently. Repeated application
of a strategy will eventually lead to a probability distribution
$p_{\text{final}}$ over configurations $\{l^{(1)}\}$ with only a single
chain each.  The quantity 
$\tilde Q(C) := \langle L \rangle(p_\text{final})$
is the expected yield of $C$ with respect to the given strategy.
Of central importance is the {\it quality} $Q(C):=\sup \tilde Q(C)$,
the best possible expected length that can be achieved starting from
$C$ by means of any strategy. We abbreviate $Q(\{1^{(N)}\})$ by
$Q(N)$ \cite{even}.  Note that the quantum nature of the cluster states does not
enter the consideration.
\begin{observation}[Lower bound for globally optimal strategy] 
Starting with $N$ EPR pairs and using {type-I} fusion gates,
the globally optimal strategy yields a cluster state of
expected length 
\begin{equation*}
	Q(N) \geq \tilde Q(N_0) + \alpha (N-N_0)
\end{equation*}
for all $N>N_0$. The constants are $N_0=92, \tilde Q(N_0)=16.1061,
\alpha=0.153336$ (known as rational numbers \cite{table}).
\end{observation}
For $N\leq 2 N_0$ a desktop computer can symbolically compute the
performance of {\sc Modesty} $\tilde Q(N) \leq Q(N)$. One finds that
the above relation is valid in this case. For $N>2 N_0$ input pairs we
adopt the following strategy: first the input is divided into $k$
blocks of length $n_i$ where $N_0 \leq n_i \leq 2 N_0$ and {\sc
Modesty} is used to convert any such block into a single chain.
Secondly, the resulting chains are fused together. 

If $C$ is a configuration consisting of only two chains of length
$l_1\geq l_2$ one easily finds that 
$Q(C)=l_1 + l_2 -2\sum_{i=0}^{l_2} 2^{-i}\geq l_1+l_2-2$. More generally, it can be shown
\cite{longpaper} that $Q(N)\geq \sum_i Q(n_i) - 2(k-1)$. Now set
$\alpha:=(\tilde Q(N_0)-2)/N_0$. From the computed data we know that
$(\tilde Q(n_i)-2)/n_i)\geq \alpha$ for all $i$. Imposing
without loss of generality $n_1=N_0$ we see that
\begin{eqnarray*}
	Q(N)
	&\geq& \tilde Q(N_0) + \sum_{i=2}^k n_i \frac{\tilde Q(n_i) -2}{n_i} \\
	&\geq& \tilde Q(N_0) + \alpha \sum_{i=2}^k n_i 
	= \tilde Q(N_0) + \alpha (N-N_0).
\end{eqnarray*}
\begin{observation}[Upper bound to globally optimal strategy]
\label{upperBound}
The quality is bounded from above by
$Q(N)\leq N/5 + 2$.
\end{observation}
While the performance of any strategy delivers a lower bound for the
optimal one, giving an {\it upper} bound is considerably harder. The
following paragraphs expose all the key ideas of a rigorous proof 
(details can be found in Ref.\ \cite{longpaper}). We proceed in 
three steps.
The first step is to realize that, because every attempted fusion
fails with probability $1/2$ and destroys two edges in case of
failure, the \emph{expected number of lost edges} equals the
\emph{expected number of fusion attempts} $T(C)$ a strategy undertakes
acting on some configuration $C$.  As the average final length $Q(C)$
is nothing other than the initial number of edges $N(C)$ minus the
expected number of losses, we have $Q(C)=N(C)-T(C)$. Hence any
\emph{lower} bound on $T$ will supply an \emph{upper} bound for
$Q(N)$.

Secondly, we pass to a greatly simplified model -- dubbed \emph{razor
model} -- from which we can extract bounds for $T$. This is done by
introducing a quite radical new rule: after every step all chains will
be cut to a maximum length of two. It turns out that there exists a
strategy in the razor model which terminates using fewer fusion
attempts on average $T_R$ than the optimal strategy for the full
model. Intuitively, this is the case as the `cutting operation'
increases the probability for chains to be completely destroyed due to
failed fusions. However, making this argument precise is greatly
impeded by the fact that one needs to compare strategies which are
defined on different models. Indeed, given the optimal strategy of the
full setup, there is no direct way of turning it into a strategy for
the razor model. We solve the problem as follows. Let $C$ be a
configuration and $C'$ the result of removing a single edge from one
chain in $C$. In \cite{longpaper} we derive the estimate $Q(C)\geq
Q(C') \geq Q(C)-1$. 
Combining the findings of the last paragraph with $N(C')=N(C)-1$, we
arrive at $Q(C')\geq Q(C) -1 \Leftrightarrow N(C) - 1 - T(C') \geq
N(C) - T(C) - 1$ and hence $T(C')\leq T(C)$. So removing a single edge
from a chain decreases the expected number of fusion attempts
performed by the optimal strategy. As the passage to the razor model
can be perceived as a repeated removal of single edges, we can use
these observations to prove $T\geq T_R$.

In a last step we further simplify the problem in order to obtain
a lower bound for $T_R$. A configuration $C$ of the razor model is
specified by two natural numbers $(l_1,l_2)$ giving the number of
chains of length $1$ and $2$, respectively. In each step a strategy has
three options: try to fuse (a) two short chains; (b) two
long ones or (c) a long and a short chain. Consider the choice
(a). In case of failure the chains are destroyed and so $C\mapsto
C+a_F$ where $a_F:=(-2,0)$. An analogous relations holds for successful
fusions where $a_S:=(-2,1)$ and similar rules can be formulated for
options $b$ and $c$. We are thus naturally led to interpret the
problem as a random walk on a two-dimensional lattice. 
As initially there are $N$ single-edge chains in the configuration,
the walk starts at $(N,0)$. It will end when there is no more than one
chain left, so at positions $(1,0), (0,1), (0,0)$. So how many steps
does a probabilistic process require -- on average -- to cover that
distance? If a strategy decides at some point in the walk to choose
action $a$, then  `on average' the configuration will move by $\bar a :=
(a_S + a_F)/2=(-2,1/2)$ on the lattice. Denote by $\langle a \rangle$
the expected number of times a given strategy opts for $a$ when acting
on $(N,0)$. Define $\bar b, \langle b \rangle, \bar c, \langle c
\rangle$ similarily. From the discussion
it is intuitive (and can be made precise \cite{longpaper}) that any
strategy fulfills $\langle a \rangle \bar a + \langle b \rangle \bar b
+ \langle c \rangle \bar c \leq (-N+1,1)$. As the expected number of
fusion attempts $T_R$ equals $\langle a \rangle + \langle b
\rangle + \langle c \rangle$, one can obtain a lower bound by solving
the \emph{linear program}: minimize $T_R$ subject to the constraints
given above. By passing to the \emph{dual problem}
\cite{Optimization}
an analytic solution can be found which gives rise to the estimate stated
in Observation \ref{upperBound}.
\begin{observation}[Symbolic calculation of optimal length]
The globally optimal strategy can be computed with
an effort of $O\big( |C_N|\,(\log|C_N|)^5\big)$.
\end{observation}
We have implemented a \emph{backtracking} algorithm which in effect
recursively computes the quality of \emph{all} configurations up to
some arbitrary total length. The results are stored in a look-up table
which causes memory consumption -- rather than time -- to limit the
practical applicability of the program. This explains the dominating
factor $|C_N|$ in the estimate of the computational effort: every
configuration has to be looked at at least once. A closer analysis
\cite{longpaper} reveals the poly-log correction. Note that, even
though the effort scales exponentially in $N$, the algorithm is vastly
more efficient than a naive approach which would enumerate all
strategies to select the optimal one by directly comparing their
performances.

The algorithm has been implemented using the computer algebra system
{\it Mathematica} and employed to derive in closed form an optimal
strategy for all configurations in $C_{46}$ \cite{table}.  A desktop
computer is capable of performing the derivation in a few hours.
Starting with $\{1^{(N)}\}$, {\sc Modesty} turns out to be the optimal
strategy for all $N\leq 10$. For configurations containing more edges,
slight deviations from {\sc Modesty} can be advantageous.  However,
the difference relative to $Q(N)$ is smaller than $1.1\times 10^{-3}$
for $N\leq 46$.
\begin{observation}[Asymptotic performance of {\sc Greed}]
Starting with $N$
EPR pairs and fusing them
with type\nobreakdash-I fusion under {\sc Greed}
results in an expected length of 
$  \tilde Q(N)= \left( 2N/\pi\right)^{1/2}
	+ O(1)$.
\end{observation}

\begin{figure}
  \includegraphics{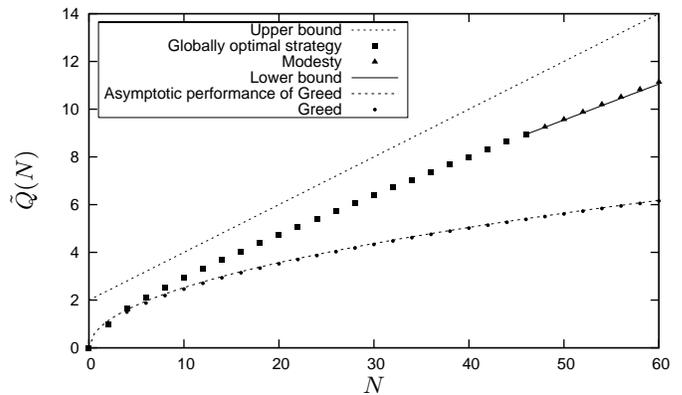}
  \caption{Expected length for the globally optimal strategy, 
  for  {\sc Modesty} (in this 
  plot indistinguishable from the former), 
  a lower bound (with $N_0=46$),
  for {\sc Greed}, its asymptotic performance, 
  and the upper bound, as
  functions of even $N$.
  \label{fig:strategies} }
\end{figure}

 It is interesting to see how {\sc Modesty} 
 compares with  the asymptotic performance of the equally reasonable
 strategy {\sc Greed}.
 Starting from  
 $\{1^{(N)}\}$, only pieces 
of length $1$ and one single piece of length $l>1$
may occur during the fusion process. 
Hence, the support of the probability distribution is 
	$\{ C= \{ l^{(1)}, 1^{(m)}\} : m=0,1,...; l=2,3,...;l+m\le N \} \cup
	\{1^{(m)}: m\le N\}$. 
	The
implementation of {\sc Greed} gives rise to a Markov
chain on this set with a reflecting boundary \cite{VK}. 
From this, one may
determine the asymptotic behaviour 
of the expected length
using a Gaussian approximation. This means the
linear chain grows as a square root in the number
of  available pairs $N$, rather than linearly.
\begin{observation}[Comparison of {\sc 
Greed} and the optimal strategy]
To realize 
an expected length of 40 in a linear cluster state, 
the resources $N$
required by {\sc Greed} and the optimal strategy already 
differ by more than an order of magnitude.
\end{observation}	
Results for the expected length using symbolic algebraic calculations
are shown in Fig.~\ref{fig:strategies}, for the strategies {\sc
Modesty}; for the globally optimal strategy, {\sc
Greed}; and the lower bound of Observation 1, 
almost identical
with the curve of {\sc Modesty}. 
The difference between the performance of {\sc Modesty}
and {\sc Greed} is enormous: it hence does matter indeed, concerning
resource consumption, what classical strategy one adopts~\cite{CNOT}.

Recall that the expected length equals the total number of edges in
the original configuration minus the expected number of losses. The
latter number, in turn, is proportional to the number of fusion
attempts on average.  Therefore, the optimum strategy is also the one
employing the smallest number of fusion steps, and is hence also the
most robust with respect to \emph{decoherence} processes associated with
these steps. Note also that the presented analysis, needless to say, can 
also be applied to other physical architectures where one has to 
cope with a probabilistic character of fusion gates, such as in 
{\it matter qubits} coupled via optical systems. 
\begin{observation}[Optimal scaling for 2-D cluster states]\label{lemma:weaving}
   An $n \times n$ cluster state can be prepared
   using a linear cluster state of length $O(n^2)$  -- employing
   $x$ measurements and type-II fusion --
   such that the overall
   success probability satisfies $P_s(n)\rightarrow 1$
   as $n\rightarrow \infty$.
\end{observation}
We now turn to two-dimensional structures, to be built by `weaving'
cluster chains. Using the type-II fusion gate~\cite{Terry} in succession
to an $x$ measurement (consuming two edges) delivers on success
($p_s=1/2$) a
vertex incorporating both linear clusters, hence an elementary 2-D
structure.
In case of failure (losing 
two edges without splitting the original
chains)
the scheme described in Ref.\ \cite{Terry} can be used for subsequent
attempts,
consuming $3+2f$ edges with $f$ being the number of failures.
Obviously, no scheme can result in more
economical asymptotics than $O(n^2)$
in the use of entangled resources.
In any preparation scheme, however, overhead has to be taken
into account to ensure a near-deterministic outcome,
as a single failure may endanger the already
generated 2-D cluster.

\begin{figure}
   \includegraphics{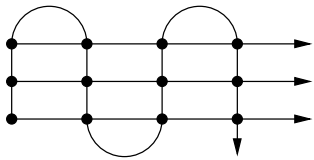}
   \caption{ A possible pattern of how to arrange
   $n+1$ linear clusters (threads) to weave a carpet of width $n$.
   Fusion operations have to be applied at the black circles
   along the long linear cluster state.
   Arrows mark free ends.
     \label{fig:2d_strategy} }
\end{figure}

Finding the overall success probability $P_s(n)$
in a closed form is impeded by the fact that
failures on earlier vertices influence the number
of resources left and therefore the number of
possible failures on later vertices.
We are able to decouple these problems by
considering a `weaving pattern'
as depicted in Fig.~\ref{fig:2d_strategy}. Let us denote with
$m$ the overhead in each of the horizonal linear
cluster states of length $l=n+m$, and
take a single linear cluster state of length
$L=n(l+1)$. We will show that a choice of
$n\mapsto a n =m$ for $a>2$ will be
an appropriate choice for the scaling of the overhead.

To start with the more formal part,
based on the above prescription,
the probability
  $P_s(n)$ of succeeding to prepare an $n\times n$
  cluster state can be written as
  $P_s(n)=      \pi_s(n )^n $. Here,
\begin{equation}\nonumber
        \pi_s(n) = \frac{1}{2^{a n}}
        \sum\limits_{k=n}^{a n } { a n \choose k} =
        1- F(n-1, a n ,1/2).
\end{equation}
is the success probability of
fusing a single chain of length $m = a n $ into the cluster,
with $F$ denoting the standard cumulative
distribution function of the binomial distribution.
Since $2n -2 \leq a n $ for all $n$,
we can hence bound $\pi_s(n)$ from below
by means of Hoeffding's inquality
\cite{Hoeff}. This gives
rise to the lower bound
        $\pi_s(n)\geq 1- \exp (
        -{2 (a n /2 - n + 1)^2 }/({a n})
        )$. As $a>2$, one can show that
        $\liminf_{n\rightarrow\infty}\pi_s(n)^n\geq 1$,
and hence, $\lim_{n\rightarrow\infty}P_s(n) =
\lim_{n\rightarrow\infty}\pi_s(n)^n =1$, which is
the argument to be shown. It is remarkable that
for $2>a>1$, then $\lim_{n\rightarrow\infty} P_S(n) =0$,
and the preparation will fail, asymptotically even with certainty.
This argument proves that a 2-D
cluster state can indeed be prepared using $O(n^2)$ EPR
pairs,
making use of probabilistic quantum gates. This may
be considered good news, as it proves that the natural
scaling in the resources can be met with negligible error.

In this work, we have addressed the 
question of how to build optical linear and two-dimensional
cluster states from the perspective of classical strategies.
We have introduced tools to assess the performance
of several protocols, including the globally optimal
strategy. Further, we have shown that
two-dimensional cluster states can be 
generated with resource requirements of 
$O(n^2)$, which is the most economical 
scaling. It has hence turned out that 
the mere classical control indeed does matter, 
and that differences in 
resource requirements of orders of magnitude
can be expected depending on the chosen strategy.
The presented techniques may, after all, 
be expected to 
provide powerful tools to assess and develop techniques
for building redundancy encoding resource states \cite{Ralph05}
or to prepare states rendering linear optical schemes
fault tolerant \cite{Trees}.

We would like to acknowledge discussions with 
G.\ Pryde, D.E.\ Browne, T.\ Rudolph,
J.\ Franson, M.B.\ Plenio, and 
O.\ Dahlsten and especially thank
A.\ Feito for key 
comments. 
This work was supported by the DFG,
the EU (QAP), the EPSRC, Microsoft 
Research, and the EURYI
scheme.


\begin{thebibliography}{99}
\bibitem{Spec}
	More specifically, it is used that it is not possible
	to discriminate unambiguously four equiprobable
	Bell states with a probability higher than $p_s=1/2$
	\cite{Lut}. The claim is that starting from EPR pairs
	and modes prepared in the vacuum, no linear optical
	fusion gate
	can operate with a higher probability of success.
\bibitem{CNOT}
	When preparing the
	appropriate graph state 
	for a single, unconnected CNOT gate \cite{Oneway,GS}
	with $P_s=0.99$,
	linear clusters of lengths $(15,12,15)$ are required.
	The difference in the number of EPR pairs used 
	between {\sc Greed} and the
	optimal strategy (resulting in these lengths on average)
	is almost a factor of $4$. In a full scheme, the
	overhead for a reliable connection would 
	additionally
	have to be included.
\end{thebibliography}
\end{document}